\documentclass{aa}

\usepackage{graphicx}
\usepackage{orcidlink}
\bibpunct{(}{)}{;}{a}{}{,}
\usepackage{gensymb}
\usepackage{float}
\usepackage{txfonts}
\usepackage{hyperref}
\usepackage{tabularx}
\hypersetup{colorlinks=true,allcolors=blue}

\begin{document} 
\renewcommand{\arraystretch}{1.5}

   \title{Resolving the terrestrial planet-forming region of HD 172555 with ALMA}
   \subtitle{I. Post-impact dust distribution}
    \authorrunning{Z. Roumeliotis et al.}
   \author{Z. Roumeliotis
          \inst{1}\fnmsep\thanks{Corresponding author; \email{roumeliz@tcd.ie}}\protect\orcidlink{0009-0003-3696-9673}
          \and L. Matr\`a \inst{1}\protect\orcidlink{0000-0003-4705-3188}
          \and G. M. Kennedy  \inst{2}\protect\orcidlink{0000-0001-6831-7547}
          \and S. Marino  \inst{3}\protect\orcidlink{0000-0002-5352-2924}
          \and K. Y. L. Su  \inst{4, 5}\protect\orcidlink{0000-0002-3532-5580}
          \and D. J. Wilner  \inst{6}\protect\orcidlink{0000-0003-1526-7587}
          \and M. C. Wyatt  \inst{7}\protect\orcidlink{0000-0001-9064-5598}
          \and A. P. Jackson  \inst{8}\protect\orcidlink{0000-0003-4393-9520}
}
   \institute{School of Physics, Trinity College Dublin, the University of Dublin, College Green, Dublin 2, Ireland
   \and Malaghan Institute of Medical Research, Gate 7, Victoria University, Kelburn Parade, Wellington 6012, New Zealand 
   \and School of Physics and Astronomy, University of Exeter, Astrophysics Group, Stocker Road, Exeter EX4 4QL, UK
   \and Space Science Institute, 4750 Walnut Street, Suite 205, Boulder, CO 80301, USA
   \and Steward Observatory, University of Arizona, 933 N. Cherry Avenue, Tucson, AZ 85721-0065, USA
   \and Center for Astrophysics, Harvard \& Smithsonian, 60 Garden Street, Cambridge, MA 02138, USA
   \and Institute of Astronomy, University of Cambridge, Madingley Road, Cambridge CB3 OHA, UK
   \and Department of Physics, Astronomy, \& Geosciences, Towson University, 8000 York Road, Towson, MD 21252, USA
             }

   \date{Received XXX / Accepted YYY}
 
  \abstract
   {Giant impacts between planetary embryos are a natural step in the terrestrial planet formation process and are expected to create disks of warm debris in the terrestrial regions of their stars. Understanding the gas and dust debris produced in giant impacts is vital for comprehending and constraining models of planetary collisions.}
   {We reveal the distribution of millimeter grains in the giant impact debris disk of HD 172555 for the first time, using new ALMA 0.87 mm observations at $\sim$80 mas (2.3 au) resolution.}
   {We modeled the interferometric visibilities to obtain basic spatial properties of the disk, and compared it to the disk's dust and gas distributions at other wavelengths.}
   {We detect the star and dust emission from an inclined disk out to $\sim$9 au and down to 2.3 au (on-sky) from the central star, with no significant asymmetry in the dust distribution. Radiative transfer modeling of the visibilities indicates the disk surface density distribution of millimeter grains most likely peaks around $\sim$5 au, while the width inferred remains model-dependent at the S/N of the data. We highlight an outward radial offset of the small grains traced by scattered light observations compared to the millimeter grains, which could be explained by the combined effect of gas drag and radiation pressure in the presence of large enough gas densities. Furthermore, SED modeling implies a size distribution slope for the millimeter grains consistent with the expectation of collisional evolution and flatter than inferred for the micron-sized grains, implying a break in the grain size distribution and confirming an overabundance of small grains.}
   {}

   \keywords{stars: individual: HD 172555 -- submillimeter: planetary systems -- techniques: interferometric -- planets and satellites: formation}

   \maketitle

\section{Introduction}

\par Even after the gas-rich protoplanetary disk phase of planetary formation ends, terrestrial planet formation continues in the inner regions of planetary systems during the era of giant impacts. This time, between $\sim$10 and 100 Myr, is dominated by the growth of planetary embryos through mutual collisions, eventually achieving final planet masses and orbital configurations \citep[e.g.][and references therein]{chambers_making_2001, morbidelli_building_2012}. There is plenty of evidence that an era of impacts took place in the Solar System, as evidenced, for example, by the formation of our own Moon \citep[e.g.][]{cameron_origin_1976, canup_origin_2001, canup_simulations_2004}, Mercury's iron enrichment \citep[e.g.][]{benz_collisional_1988, cameron_strange_1988, benz_origin_2007}, and the Martian hemispheric dichotomy \citep[e.g.][]{wilhelms1984, smith_global_1999, andrews-hanna_borealis_2008}. Outside the solar system, indirect evidence is found in exoplanet populations, for example in the statistics of mature close-in super-Earth planetary systems \citep[][and references therein]{izidoro_formation_2018}, and in the presence of warm dust in the inner few au regions of $\sim$3\% of 10-100 Myr-old stars \citep{kennedy_bright_2013}. 
\par In general, it is difficult to distinguish between steady-state phenomena (asteroid belt analogs, e.g. \citeauthor{su_asteroid_2013} \citeyear{su_asteroid_2013}) and non-steady-state phenomena (giant impacts or other transient events, e.g. \citeauthor{geiler_does_2017} \citeyear{geiler_does_2017}) as the origin of the warm dust in the inner few au regions of these stars. In an effort to distinguish between disks created by steady-state and non-steady-state phenomena, \citet{wyatt_transience_2007} created an analytical model for the steady-state collisional evolution of disks which shows that, at a given radial location and system age, there is a maximum possible fractional infrared luminosity a disk can have due to collisional processing. Disks with an infrared luminosity below this maximum can be explained by steady-state phenomena, such as asteroid belts, but disks with infrared luminosities above the maximum allowed value cannot be produced in this way and instead must be undergoing a transient event, such as a giant impact \citep{wyatt_transience_2007}. 
In a few of these extremely high fractional luminosity systems, mid-infrared spectroscopy additionally shows a unique dust mineralogy with features from glassy silica, a thermodynamically altered mineral that requires high temperature processing and vapor condensation, naturally provided by a hypervelocity, planetary-scale impact \citep[e.g.][]{lisse_abundant_2009, johnson_self-consistent_2012}. Of the known young, warm, and extremely dusty debris disks, only three systems in the expected age of terrestrial planet formation show this glassy silica feature in their \textit{Spitzer} spectra: HD 23514 \citep{rhee_warm_2008, meng_variability_2012}, HD 15407A \citep{fujiwara_silica-rich_2012, olofsson_transient_2012}, and the focus of this paper, HD 172555 \citep{chen_spitzer_2006, lisse_abundant_2009}. Recent \textit{James Webb} Space Telescope (JWST) mid-infrared spectra further reveal a handful of newly identified extremely dusty debris disks that are rich in silica dust (Su et al. in prep.). 
\par With multiple lines of evidence pointing to a giant impact scenario as the origin of the debris, HD 172555 (HR 7012, HIP 92024) is one of the best characterized systems with warm dust. HD 172555 is an A7V \citep{gray_contributions_2006} star at a distance of 28.79$\pm$0.13 pc from Earth \citep{gaia_collaboration_vizier_2022} and is a member of the $\beta$ Pictoris moving group \citep{zuckerman__2001}. The age of the moving group has been estimated to be 23.4$\pm$4.8 Myr \citep[][]{olivares_bayesian_2025}. HD 172555 hosts a close to edge-on debris disk, with an inclination of 76.2\degree$\pm$1.7\degree, reported by \citet{engler_detection_2018} as 103.8\degree$\pm$1.7\degree. The system's infrared fractional luminosity is $7.2\times10^{-4}$ \citep{mittal2015}, which is $\sim$300 times larger than the maximum allowed infrared luminosity for the age and disk radius of HD 172555 \citep{wyatt_transience_2007}. SED fitting of the system shows the infrared excess is best fit by warm, $\sim$290K dust \citep{cote_b_1987}, and fitting of the mid-infrared \textit{Spitzer} continuum and solid-state features requires a non-steady-state particle size distribution, with an overabundance of small grains \citep{lisse_abundant_2009, johnson_self-consistent_2012}. The latter also supports the impact scenario, as the overabundance of small grains indicates that the circumstellar material must have been created relatively recently and in a transient event.
\par The final piece of evidence for the impact scenario comes from the Atacama Large Millimeter/submillimeter Array (ALMA) detection of \textsuperscript{12}CO J=2-1 emission at 1.3 mm \citep{schneiderman_carbon_2021}. These Cycle 1 observations detect both the continuum and \textsuperscript{12}CO, which are both spatially unresolved, with a resolution of 1.16"$\times$0.75" (33.4$\times$21.6 au). However, using the fact that the \textsuperscript{12}CO is spectrally resolved, \citet{schneiderman_carbon_2021} constrain the CO to a ring of radius $\sim$7.5 au. At this location around an A star, asteroids would be too warm to retain CO or CO\textsubscript{2} ice for subsequent release over the age of the system \citep[e.g.][and references therein]{prialnik_can_2009, snodgrass_main_2017}. \citet{schneiderman_carbon_2021} explore four scenarios for the origin of both the dust and CO in the system: leftover from a primordial protoplanetary disk, collisional production in an asteroid belt, inward transport from an outer reservoir, and release in the aftermath of a giant impact. They favor the giant impact between planetary-sized bodies with atmospheric stripping scenario as it is the only one that can explain the dust mineralogy, particle size distribution, dust mass, radial distribution of dust and CO, and total amount of CO detected. However, because these observations were spatially unresolved, the spatial distribution of the millimeter grains remains unknown. 
\par Observations of HD 172555 at other wavelengths reveal the presence of smaller dust grains and atomic gas in the system. In the mid-infrared, using multiepoch photometry and spectroscopy, no evidence was found for variability in the HD 172555 system \citep{su_mid-infrared_2020}. Combined with later JWST observations, this indicates that the submicron grains produced in the impact are stable on decades-long timescales \citep{su_mid-infrared_2020, samland_minds_2025}. HD 172555 has been spatially resolved in the mid-infrared using interferometry and imaging \citep{smith_resolving_2012} and in the optical using scattered light imaging \citep{engler_detection_2018}; both of these observations constrain the outer edge of the dust emission to 8-10 au and present marginal evidence for an asymmetry in the dust distribution. Overall, the small grains probed by the mid-infrared and scattered light observations appear to be roughly co-located with the CO detected by ALMA \citep{schneiderman_carbon_2021}, suggesting that the impact took place in this region. 
\par The innermost region of the HD 172555 planetary system (a few stellar radii to $<$0.5 au) shows signs of inward transport of material and dynamic upheaval. Star-grazing bodies (exocomets) have been detected transiting at $\sim$7 stellar radii from the central star through time-variable absorption in the Ca II K and H doublet lines in the UV and in the optical light curve \citep{kiefer_exocomets_2014, kiefer_hint_2023}. This provides further evidence for the planetary system being close to edge-on. \citet{kiefer_hint_2023} calculated the evaporation efficiency for both the exocomet transit they found in the optical light curve and the exocomets detected spectroscopically by \citet{kiefer_exocomets_2014} and concluded there are likely at least two classes of exocomets in this system. Some of the gas emission in this system has been attributed to these exocomets. Using the \textit{Hubble} Space Telescope, \citet{grady_star-grazing_2018} detected \ion{Si}{III}, \ion{Si}{IV}, \ion{C}{II}, \ion{C}{IV}, and \ion{O}{I} absorption from exocomets originating from the warm dust disk and potentially perturbed onto star-grazing orbits by a Jovian-mass planet. \ion{Cl}{I}, \ion{S}{I}, \ion{Ni}{II}, and \ion{Fe}{II} emission from a gaseous disk $<$0.5 au from the star, believed to be from evaporating rocky bodies, was detected with JWST \citep{samland_minds_2025}. This hot gas is likely to have been predominantly released in-situ and could be linked to the giant impact, as the impact could have led to increased dynamical activity in the system which led to increased collisional interactions or the production of dust that can drift inward and sublimate \citep[][and references therein]{samland_minds_2025}. Finally, spectrally and spatially unresolved \ion{O}{I} emission from a circumstellar disk was detected with the \textit{Herschel} Space Observatory, which could have been released in the giant impact or accumulated slowly over time from collisions within a belt of dust \citep{riviere-marichalar_hd_2012}. The \ion{O}{I} emission could have originated in the giant impact, as silicate constituents (Si, Fe, Mg, O) can be produced in gas form during a hypervelocity impact \citep{pahlevanChemicalFractionationSilicate2011}, or from the photodissociation of volatile molecules originating in the stripped atmosphere \citep{schneiderman_carbon_2021}. However, we do not presently know whether the \ion{O}{I} is co-located with the CO or with the hot atomic gas revealed by JWST.
\par In order to better constrain the radial and azimuthal distribution of the impact-produced dust at a few au around HD 172555, we present follow-up high resolution 0.87 mm continuum observations with ALMA. In Sect. \ref{sect:observations and data prep} we describe our observations, calibration, and imaging. In Sect. \ref{sect:results}, we present the first resolved image of dust around HD 172555 at millimeter wavelengths. Section \ref{sect:modeling} describes the modeling we performed on the data visibilities to determine basic disk parameters. In Sect. \ref{sect:discussion} we discuss our findings in the context of the literature, and conclude with a summary in Sect. \ref{sect:summary}.

\section{Observations}  
\label{sect:observations and data prep}
\par We observed the HD 172555 system for a total of 6.6 hours on-source with ALMA on Chajnantor, Chile during Cycle 9 through project 2022.1.00793.S (PI: Matr\`a). Observations were split between a lower and higher spectral setup covering different frequencies within Band 7 (0.87 mm). All observations were carried out using Band 7 receivers and the 12-m array with baselines ranging between 27.5 and 3637.7 m for each setup. Observations in the lower spectral setup were taken on five separate days between 16 May and 3 June 2023, and observations in the higher spectral setup were taken at four times between 3 and 4 June 2023. 
\par In each of the higher and lower spectral setups, four unique spectral windows were used. In the higher setup, all four of the spectral windows were 1.875 GHz wide, centered at 343.2, 345.2, 355.2, and 357.2 GHz, with a channel width of 488.281 kHz. In the lower setup, three of the spectral windows were 1.875 GHz wide, centered at 328.9, 330.9, and 340.9 GHz, and one spectral window was 2 GHz wide centered at 342.9 GHz, with a resolution of 31.3 MHz. The windows in the higher spectral setup were set to cover the CS J=7-6, \textsuperscript{12}CO J=3-2, HCN J=4-3, and HCO+ J=4-3 transitions at 342.883, 345.796, 354.505, and 356.734 GHz, respectively. The windows of the lower spectral setup were set to cover the C\textsuperscript{18}O J=3-2, \textsuperscript{13}CO J=3-2, and CN J=3-2 transitions at 329.331, 330.588, and 340.248 GHz, respectively, as well as the continuum. 
\par Standard calibrations were applied to each visibility dataset by the ALMA observatory using its pipeline. Data manipulation was carried out using the Common Astronomy Software Applications (CASA) software version 6.6.5 \citep{the_casa_team_casa_2022}. For the continuum analysis, for each of the four spectral windows in the two spectral setups, we flagged the spectral lines, then time and frequency averaged the visibilities (to 30 seconds and 1.875 GHz, the size of the entire spectral window, respectively) to facilitate the processing of the large dataset while avoiding time- or frequency-smearing effects. We then concatenated all datasets for each spectral window to obtain a final calibrated and time- and frequency-averaged continuum visibility dataset for each spectral setup. We did not implement any astrometric realignment for each observation, as the proper motion and parallax of the system cause motions of at most $<$12 mas over the maximum time difference between observations, which are about an order of magnitude smaller than the beam and comparable to to the absolute astrometric accuracy\footnote{\url{https://help.almascience.org/kb/articles/what-is-the-absolute-astrometric-accuracy-of-alma}} of our continuum data, $\sim$13 mas. 
\par We jointly imaged the higher and lower spectral setups in CASA using the CLEAN algorithm \citep{hogbom_aperture_1974} implemented through the tclean task. We carried out the continuum imaging in multi-frequency synthesis mode with Hogbom deconvolution \citep{hogbom_aperture_1974}. We chose a natural weighting scheme with a u-v taper of 0.08" to achieve maximum surface brightness sensitivity (S/N beam$^{-1}$) while still maintaining some spatial resolution, which led to a beam of 166$\times$148 mas (4.8$\times$4.3 au at the distance of HD 172555) and an RMS noise level of 10.4 µJy beam$^{-1}$. At the same time, we also created an image using Briggs weighting \citep{briggs_high_1995} with a standard robust value of 0.5, obtaining a resolution of 84$\times$73 mas (2.4$\times$2.1 au) with an RMS noise level of 11.3 µJy beam$^{-1}$. 
\par The primary focus of this work is the analysis of the dust continuum emission, with in-depth analysis of the spectral lines deferred to a future companion paper covering their detection, kinematics, and compositional analysis. 
\section{Results}
\label{sect:results}
\begin{figure*}[!h]
    \sidecaption
    \centering
    \includegraphics[width=12cm]{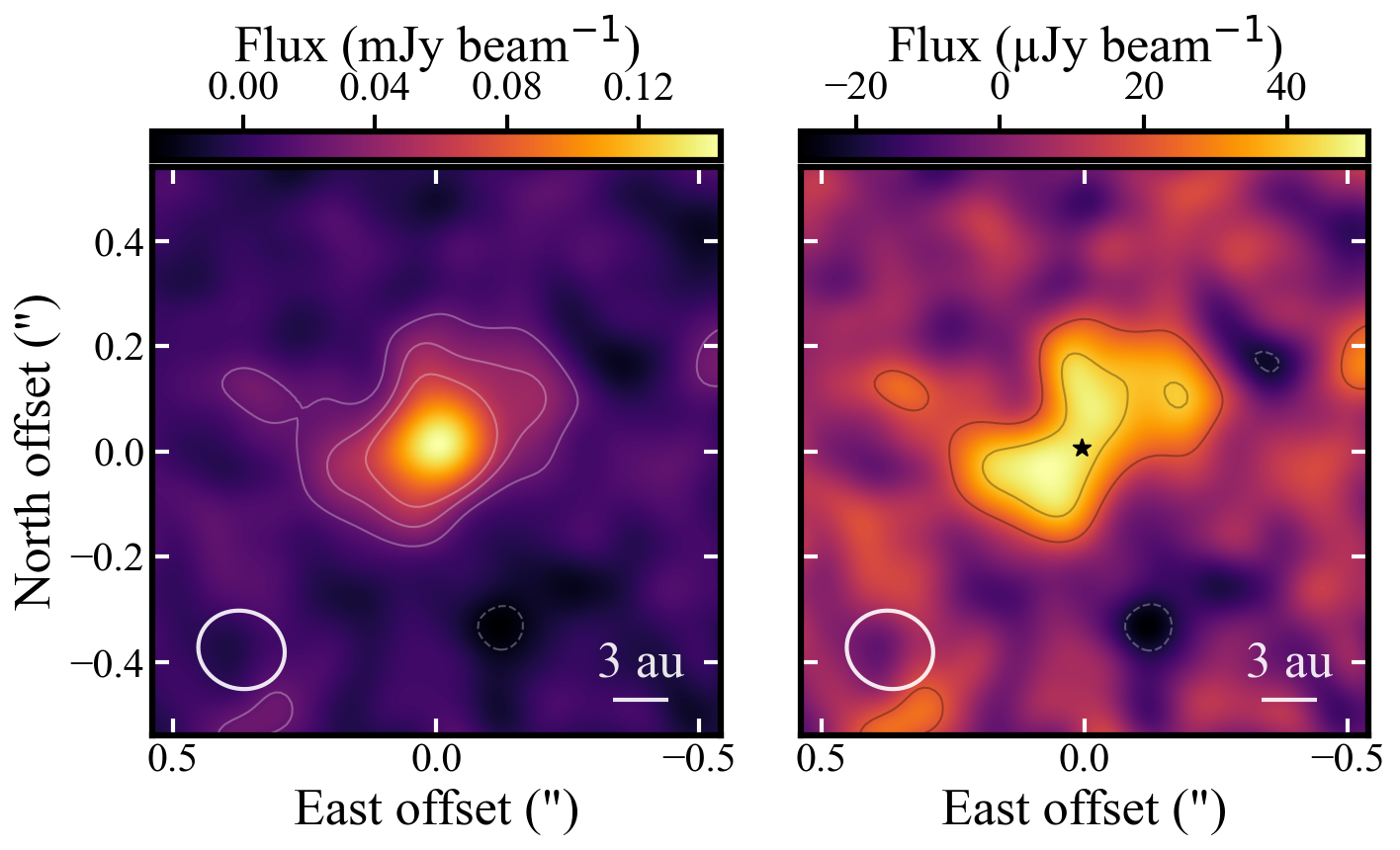}
    \caption{Left: ALMA 0.87 mm continuum emission from the HD 172555 planetary system (star and disk) obtained by joint imaging of the combined visibility dataset. Right: same as the left panel, but imaged after removing the star from the visibilities, as described in Sect. \ref{sect:results}. The symbol in the center notes the position of the star. In both panels, north is up and east is left. Contours are $\pm$[2, 4, 6] $\times$ 10.4 µJy beam$^{-1}$, the RMS noise level. Images are made with a natural weighting and a u-v taper, as described in Sect. \ref{sect:observations and data prep}.}
    \label{fig:all data plus diskonly}
\end{figure*}
\par The left panel in Fig. \ref{fig:all data plus diskonly} shows the final continuum (tapered) image with the combined thermal emission from the central star and the disk in the HD 172555 system. In order to better constrain the morphology of the circumstellar dust without contamination from the central star, we aim to create a disk-only continuum image using a procedure similar to \citet{matra_dust_2020}. In order to do this, we assume the star is a point source and appears with constant amplitude on long baselines. We then iteratively image the long-baseline visibilities with a minimum u-v distance cutoff, which we increase at every iteration. We visually identify the minimum u-v distance which allows us to remove the extended disk emission while still detecting the compact, unresolved central star. We find the best compromise to be a minimum u-v distance of 600 k$\lambda$ for the lower spectral setup and 700 k$\lambda$ for the higher spectral setup. 

\par Separately for each spectral setup, assuming the star is an unresolved point source, we fit the visibilities of the longest baselines with the above u-v cutoffs, now unbiased by any disk emission, to obtain a best-fit stellar flux and RA/Dec position. The results of our stellar fits for each spectral setup are detailed in Table \ref{tab:star only mcmc values}. We note that the fluxes are consistent within the errors, which include both random noise and flux calibration uncertainty. We then subtract the best-fit model visibilities for each spectral setup from the entire visibility dataset for that setup, including all baselines, effectively subtracting the star from our data. Finally, we imaged the star-subtracted visibilities, combining both spectral setups, to obtain a disk-only continuum image, shown in the right panel of Fig. \ref{fig:all data plus diskonly}.

\begin{table}
\caption{Results from fitting the stellar flux and position.}
\centering
\small
\begin{tabularx}{\hsize}{lll}
\hline
Parameter  & Lower Spectral Setup & Higher Spectral Setup \\
\hline
Stellar Flux (mJy)  & $0.10\pm0.03$ & $0.07\pm0.03$  \\
RA offset (mas)                     & $3.7\pm9.2$  & $7^{+14}_{-12}$ \\
Dec offset (mas)                    & $10.1^{+8.7}_{-7.8}$ & $0\pm11$ \\
\hline
\end{tabularx}
\label{tab:star only mcmc values}
\end{table}

\par As seen in both panels of Fig. \ref{fig:all data plus diskonly}, extended emission from a close to edge-on dust disk is clearly detected along a position angle of $\sim$120º, extending out to $\sim$0.3", which corresponds to $\sim$8.6 au from the central star at the distance of HD 172555. The emission appears broad, extending down to the inner resolution element of our observations. Even after accounting for the disk inclination, this indicates dust is emitting down to 2.3 au on-sky from the central star. The orientation and outer radius of the disk are qualitatively consistent with previous scattered light and mid-infrared imaging \citep{smith_resolving_2012, engler_detection_2018}, although our observations are tracing larger millimeter grains, as opposed to the smaller grains traced by the scattered light and mid-infrared observations. Our ALMA observations reveal the inner $<$5 au region free from contamination from the central star, as well as spatially resolve the disk continuum in the millimeter, for the first time. 
\par To compare the results of our observations to what would be expected given the flux density measured by \citet{schneiderman_carbon_2021} and the spectral slope of this system in the millimeter, we measure the flux density of the disk from our disk-only images. Using the CASA task imview and visually defining a region around the disk, we measure the flux density from images created separately from the visibilities from each spectral setup, as well as from imaging of both spectral setups together, as seen in Fig. \ref{fig:all data plus diskonly}. For the lower ($\lambda_{\rm c}=0.89$ mm) and higher ($\lambda_{\rm c}=0.86$ mm) spectral setups we find $F_{\rm \nu}=0.26\pm 0.04$ mJy and $F_{\rm \nu}=0.17\pm 0.03$ mJy, respectively, and when we image the two spectral setups together, as seen in the right panel of Fig. \ref{fig:all data plus diskonly}, we measure $F_{\rm \nu}=0.18\pm 0.02$ mJy, where all quoted errors are 1$\sigma$. After accounting for the expected stellar contribution, \citet{schneiderman_carbon_2021} report $F_{\rm \nu}=0.09\pm 0.03$ mJy. Given the observed emission at 1.3 mm by \citet{schneiderman_carbon_2021} and the best-fit millimeter spectral slope (see Sect. \ref{sect:sed}), the emission we observe is consistent with the expected emission within 2$\sigma$. 

\section{Modeling}
\label{sect:modeling}

\subsection{The physical model}
\label{sect:physical model}
\par In order to formally constrain the radial structure and geometry of dust, we fit our visibilities using two models, each with two components: an axisymmetric dust disk, and the star modeled as a point source. In our first model (hereafter Gaussian model), the radial surface mass density of the disk $\Sigma$ is modeled as a Gaussian, used to determine a centroid radius $R$ and full width at half maximum (FWHM, hereafter width) $\Delta R$. For simplicity, we model the vertical density distribution of the disk as a Gaussian with the aspect ratio, $h = \frac{H}{r}$, radially fixed to 0.05, as the disk is not vertically resolved. The full dust density distribution in the disk is thus represented as 
\begin{equation}
    \rho(r,z) = \Sigma_{\rm scale} e^{-\frac{(r-R)^2}{2\sigma_{\rm r}^2}}\frac{1}{\sqrt{2\pi}\sigma_{\rm z}}e^{-\frac{z^2}{2\sigma_{\rm z}}},
    \label{eqn:surface density distribution}
\end{equation}
where $\Sigma_{\rm scale}$ is a normalization factor that is proportional to the disk flux, where we fit for the latter (see Sect. \ref{sect:modeling process}), $r$ and $z$ are the cylindrical coordinates, $R$ and $\sigma_{\rm r}$ are the center and standard deviation of the radial Gaussian disk, respectively, and $\sigma_{\rm z} = H = hr$. $\sigma_{\rm r}$ and $\Delta R$ are related through $\sigma_{\rm r} = \frac{\Delta R}{2 \sqrt{2\ln2}}$. 
\par In our second model (hereafter power law model), the radial surface mass density of the disk $\Sigma$ is modeled as a power law, which we use to determine an inner radius $R_{\rm in}$ and outer radius $R_{\rm out}$. The vertical density distribution is again modeled as a Gaussian with a constant aspect ratio of 0.05. The full dust density distribution in the disk for the power law model is represented as
\begin{equation}
    \rho(r,z) = \Sigma_{\rm scale} (\frac{r}{R_{\rm in}})^p\frac{1}{\sqrt{2\pi}\sigma_{\rm z}}e^{-\frac{z^2}{2\sigma_{\rm z}}},
    \label{eqn:surface density distribution power law}
\end{equation}
where $p$ is the power law exponent and all other variables have the same meaning as in Eq. \ref{eqn:surface density distribution}. Where $r<R_{\rm in}$ or $r>R_{\rm out}$, $\rho(r,z)=0$.
\par For both models, we set the radial temperature dependence by assuming the grains act as blackbodies around a star of 7.7 L$_\odot$, with the temperature proportional to $r^{-1/2}$. We note that the retrieved surface density distributions we will discuss are  dependent on our choice of radial temperature profile. We fix the opacity of the grains and $\Sigma_{\rm scale}$ to low enough values to ensure the disk is optically thin, as is expected, as we instead fit for the total disk flux (see Sect. \ref{sect:modeling process}).

\subsection{The visibility modeling process}
\label{sect:modeling process}
\par Separately for each of the physicals model described in Sect. \ref{sect:physical model}, we used the RADMC-3D\footnote{\url{https://www.ita.uni-heidelberg.de/~dullemond/software/radmc-3d/}} \citep{dullemond_radmc-3d_2012} radiative transfer code to create an image of the disk at 0.87 mm. Initially, the disk is centered at the image origin and is inclined from the plane of the sky by an inclination angle $i$ and rotated so that its semimajor axis is at a position angle PA compared to the direction of declination, measured east of north; both $i$ and PA are free parameters. For the Gaussian model, $R$ and $\Delta R$, defined in Sect. \ref{sect:physical model}, are also both free parameters. For the power law model, the free parameters $R_{\rm in}$ and $R_{\rm out}$, defined in Sect. \ref{sect:physical model}, are used instead of $R$ and $\Delta R$, and we additionally include the power law exponent $p$ as a free parameter. All other free parameters are the same in the two models. In both models, the specific intensity in the disk-only image is normalized, then rescaled so that the integral of the pixel surface brightness over the image is equal to the model disk's flux density. We separately normalize and rescale the higher and lower spectral setups, leaving the flux density in each of the setups, $F^{\rm higher}_{\mathrm{\nu_{d}}}$ and $F^{\rm lower}_{\mathrm{\nu_{d}}}$, as free parameters. Both spectral setups use the same values of $i$, PA, and either $R$ and $\Delta R$ (Gaussian model) or $R_{\rm in}$, $R_{\rm out}$, and $p$ (power law model).
\par We use the GALARIO\footnote{\url{https://github.com/mtazzari/galario/}} software package \citep{tazzari_galario_2018} to obtain a Fourier transform of the model image from RADMC-3D and sample it at the same u-v locations as the data from the higher and lower spectral setups. We add the star to the model visibility as a point source component with flux density $F^{\rm higher}_{\mathrm{\nu_{*}}}$ and $F^{\rm lower}_{\mathrm{\nu_{*}}}$, both of which are free parameters, for the respective datasets. Finally, we apply an RA and Dec offset ($\Delta$RA and $\Delta$Dec, respectively), each of which is left as a free parameter, to the entire model (star and disk) as a phase shift in Fourier space. Both of the spectral setups use the same $\Delta$RA and $\Delta$Dec. 
\par Our final free parameter is a rescaling factor $f$ for the weight of each data point. The weight on each u-v point contains the uncertainty $\sigma$ on the real and imaginary components of the data point. The weight and uncertainty are related by $w=1/\sigma^2$, and the weights are delivered by the observatory as calculated in the calibration process. However, it has been shown that these weights, while accurate for different u-v points relative to one another within a dataset, can be inaccurate between datasets, and need to be rescaled by a factor common to all visibilities within a given dataset \citep[e.g.][]{marino_gap_2018, matra_kuiper_2019}.

\begin{figure*}[!h]
    \centering
    \includegraphics[width=\hsize]{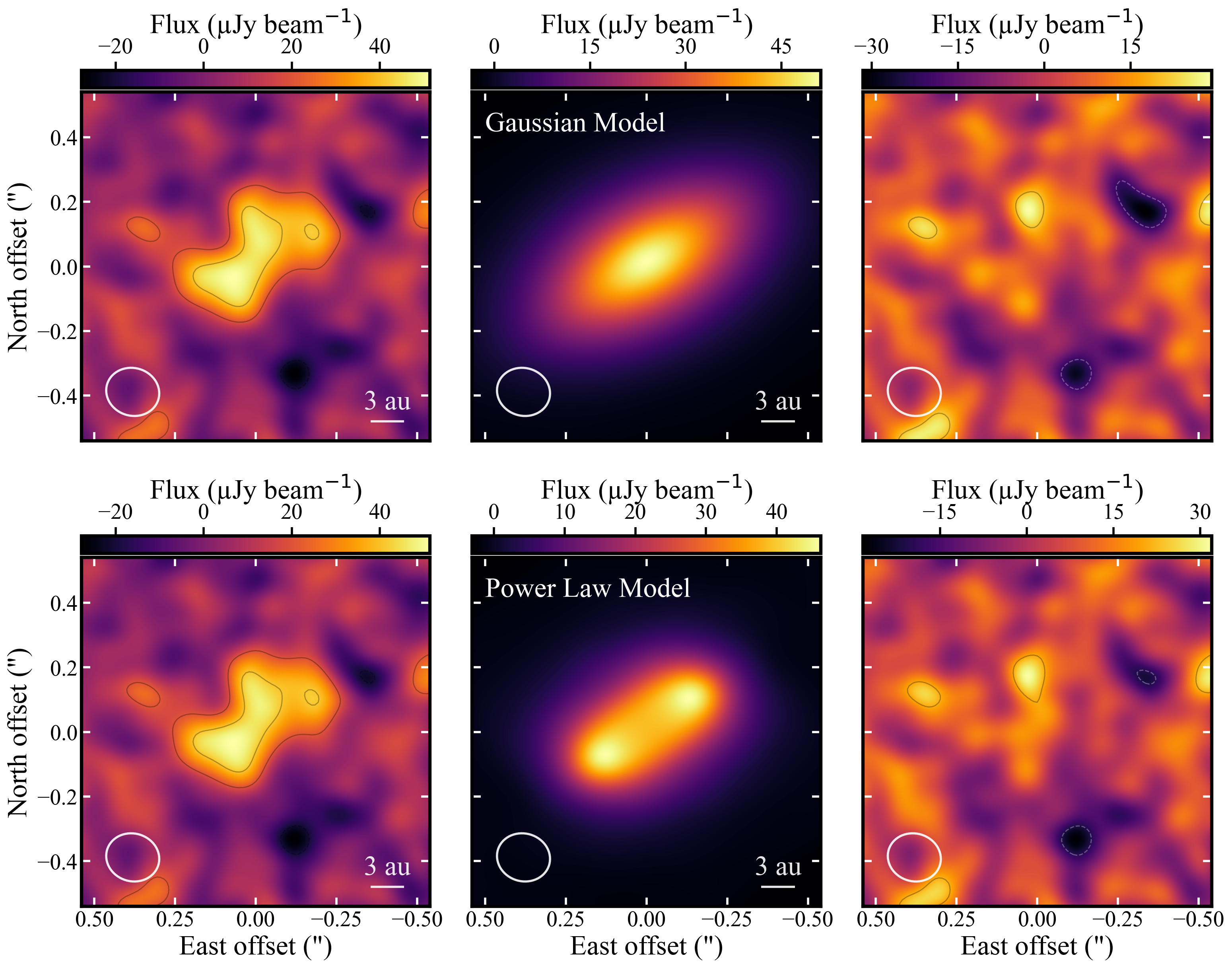}
    \caption{Left column: same as the right panel in Fig. \ref{fig:all data plus diskonly}. Middle column: disk-only models created using RADMC-3D and tclean with the best-fit values from the MCMC fitting (see Table \ref{tab:final values}), as described in Sect. \ref{sect:modeling process}. Right column: residuals from subtracting the model visibilities from the data visibilities, as described in Sect. \ref{sect:modeling results}. Top row: Gaussian model and residuals corresponding to that model. Bottom row: power law model and residuals corresponding to that model. All panels are imaged using a natural weighting and a u-v taper, as described in Sect. \ref{sect:observations and data prep}, and the contours are $\pm$[2, 4, 6] $\times$ 10.4 µJy beam$^{-1}$, the RMS noise level.}
    \label{fig:data model residuals}
\end{figure*}

\par We fit the model visibilities in both the higher and lower spectral setups simultaneously using the affine-invariant Markov Chain Monte Carlo (MCMC) ensemble sampler from \citet{goodman_ensemble_2010}, implemented through the EMCEE v3.1.6 software package \citep{foreman-mackey_emcee_2013, foreman-mackey_emcee_2019}. The likelihood function is proportional to $e^{-\chi_{\rm total}^2/2}$, where $\chi_{\rm total}^2$ is the sum of the two $\chi^2$ from the fits to the higher and lower setups. Initially, for all of the model parameters, we used uniform priors, with ranges chosen to retain physical significance while allowing the chains to explore a wide region of parameter space. We ran the MCMC to sample the posterior probability distribution of the parameters using 110 walkers (10 times the number of free parameters), and for 10000 steps. We ensured visual convergence of the MCMC chains, and repeated this procedure for both models described in Sect. \ref{sect:physical model}.
\par After running the MCMC for the Gaussian model with uniform priors on all of the parameters, we noted that the inclination was not well constrained, so we placed a Gaussian prior, with a mean of 76.5\degree and standard deviation of 8.2\degree, only on the inclination using the results from modeling the inclination done by \cite{engler_detection_2018} and ran the MCMC again with 110 walkers for 10000 steps. The remaining parameters retained the same uniform priors. The MCMC for the power law model used the same Gaussian prior on the inclination and otherwise uniform priors on the remaining parameters.

\subsection{Visibility modeling results}
\label{sect:modeling results}
\par For most of the parameters, Table \ref{tab:final values} reports the $50^{+34}_{-34}$th percentile values of the posterior probability distribution of each parameter marginalized over all other parameters for both of the models described in Sect. \ref{sect:physical model} after discarding the burn-in phase of 400 steps. Figures \ref{fig:cornerplot incl gaussian prior 10000 steps} and \ref{fig:cornerplot incl gaussian prior 10000 steps power law} show the posterior probability distributions for each model. We note that for $R_{\rm in}$ and $R_{\rm out}$, Table \ref{tab:final values} reports the mode (highest probability value) and 68\% highest density interval instead of the $50^{+34}_{-34}$th percentile values of the posterior probability distributions, as their joint two-dimensional probability distribution is significantly asymmetric, making the median (50th percentile) an inaccurate representation of the joint highest probability values (see Fig. \ref{fig:cornerplot incl gaussian prior 10000 steps power law}).

\begin{table}[h]
\caption{Results from MCMC modeling.}
\centering
\begin{tabular}{lll}
\hline
Free Parameter              & Gaussian Model & Power Law Model                         \\
\hline
$F^{\rm lower}_{\mathrm{\nu_{*}}}$ (mJy)  &  $0.10^{+0.03}_{-0.03}$ & $0.12^{+0.02}_{-0.02}$\\
$F^{\rm higher}_{\mathrm{\nu_{*}}}$ (mJy) &  $0.08^{+0.02}_{-0.03}$ & $0.09^{+0.02}_{-0.02}$\\
$F^{\rm lower}_{\mathrm{\nu_{d}}}$ (mJy)  &  $0.26^{+0.09}_{-0.08}$ & $0.19^{+0.06}_{-0.06}$\\
$F^{\rm higher}_{\mathrm{\nu_{d}}}$ (mJy) &  $0.18^{+0.08}_{-0.08}$ & $0.13^{+0.06}_{-0.06}$\\
$\Delta$RA (mas)                     &  $5.5^{+6.2}_{-6.1}$ & $5.6^{+5.9}_{-6.1}$\\
$\Delta$Dec (mas)                    &  $6.3^{+5.8}_{-5.7}$ & $6.9^{+5.8}_{-5.5}$\\
PA (º)                      & $121.5^{+23.1}_{-18.7}$  & $123.3^{+25.0}_{-14.7}$ \\
$i$ (º)                     & $71.5^{+6.3}_{-6.4}$ & $71.9^{+6.8}_{-6.6}$\\
$R$ (au)                      & $4.7^{+2.8}_{-2.4}$  & -- \\
$\Delta R$ (au)              & $12.9^{+4.8}_{-5.7}$ & -- \\
$R_{\rm in}$ (au) & -- & $4.7^{+0.3}_{-2.1}$\\
$R_{\rm out}$ (au) & -- & $7.1^{+3.3}_{-1.7}$\\
$p$ & -- & $0.0^{+7.9}_{-5.1}$\\
$f$                           & $0.1437^{+0.0001}_{-0.0001}$ & $0.1437^{+0.0001}_{-0.0001}$  \\
\hline
\end{tabular}
\tablefoot{Middle column: results from Gaussian model. Right column: results from power law model. In both columns, the inclination $i$ has a Gaussian prior applied (see Sect. \ref{sect:modeling process}).}
\label{tab:final values}
\end{table}

\par From Table \ref{tab:final values}, in both models, the stellar position ($\Delta$RA and $\Delta$Dec) is consistent with the phase and image center of our observations. The PA and $i$ in both models are consistent with what we expected qualitatively from the images, although they are uncertain at the relatively low S/N of our observations. The modeled disk and stellar fluxes are consistent between both models. The results for $R$ and $\Delta R$ show there is a preference in the Gaussian model for broad disks centered around $\sim$5 au, although there is a degeneracy between $R$ and $\Delta R$ in our fit (see Fig. \ref{fig:cornerplot incl gaussian prior 10000 steps}). On the other hand, there is a preference in the power law model for narrower disks between $\sim$5 and $\sim$7 au, although we note that $R_{\rm in}$ is not constrained (see Fig. \ref{fig:cornerplot incl gaussian prior 10000 steps power law}). We also note that there is a degeneracy between both $p$ and $R_{\rm in}$ and $p$ and $R_{\rm out}$, but $p$ is most likely $\sim$0, corresponding to a relatively flat surface density between $R_{\rm in}$ and $R_{\rm out}$ (see Fig. \ref{fig:cornerplot incl gaussian prior 10000 steps power law}).
\par To assess the goodness of fit, we produced a model image with RADMC-3D using the best-fit parameters from modeling. For the most part, this is the median, but, as discussed above, we use the mode for both $R_{\rm in}$ and $R_{\rm out}$ in the power law model. Then, separately for each spectral setup, we created visibilities from that model so that we could image the visibilities from both spectral setups together. We created a dirty image of the joint visibility model dataset, using the same imaging parameters described in Sect. \ref{sect:observations and data prep}, seen in the center column in Fig. \ref{fig:data model residuals}. The central brightness peak seen in the middle of the Gaussian model image (Fig. \ref{fig:data model residuals}, top center), is due to the steep blackbody temperature distribution, combined with the Gaussian best-fit model dust density profile being broader and extending significantly closer to the central star. Finally, to evaluate the goodness of fit, we subtracted the model visibilities from the data visibilities separately for each spectral setup to produce residual visibilities. We then produced a joint residual dirty image (right column in Fig. \ref{fig:data model residuals}) using the same imaging parameters as the data. We repeated this procedure for each model described in Sect. \ref{sect:physical model} (rows in Fig. \ref{fig:data model residuals}). The residual dirty images produced for each model show a lack of significant residuals, indicating that both models are a good fit to the ALMA data and that our data cannot distinguish between the two models.

\subsection{SED modeling}
\label{sect:sed}
\par From our disk-only images, we find $F^{\rm lower}_{\mathrm{\nu_{d}}}=0.26\pm0.04$ mJy and $F^{\rm higher}_{\mathrm{\nu_{d}}}=0.17\pm0.03$ mJy (see Sect. \ref{sect:results} for more details). From modeling the data visibilities (Sect. \ref{sect:modeling results}), we find $F^{\rm lower}_{\mathrm{\nu_{d}}}=0.26^{+0.09}_{-0.08}$ mJy and $F^{\rm higher}_{\mathrm{\nu_{d}}}=0.18^{+0.09}_{-0.08}$ mJy. We note that these fluxes are fit by the Gaussian model, but the fluxes are consistent between both models within $1\sigma$. The flux densities from the visibilities and from the images are consistent within each spectral setup, which implies that we are not missing flux on large scales. 

\par
\begin{figure*}
    \centering
    \includegraphics[width=\hsize]{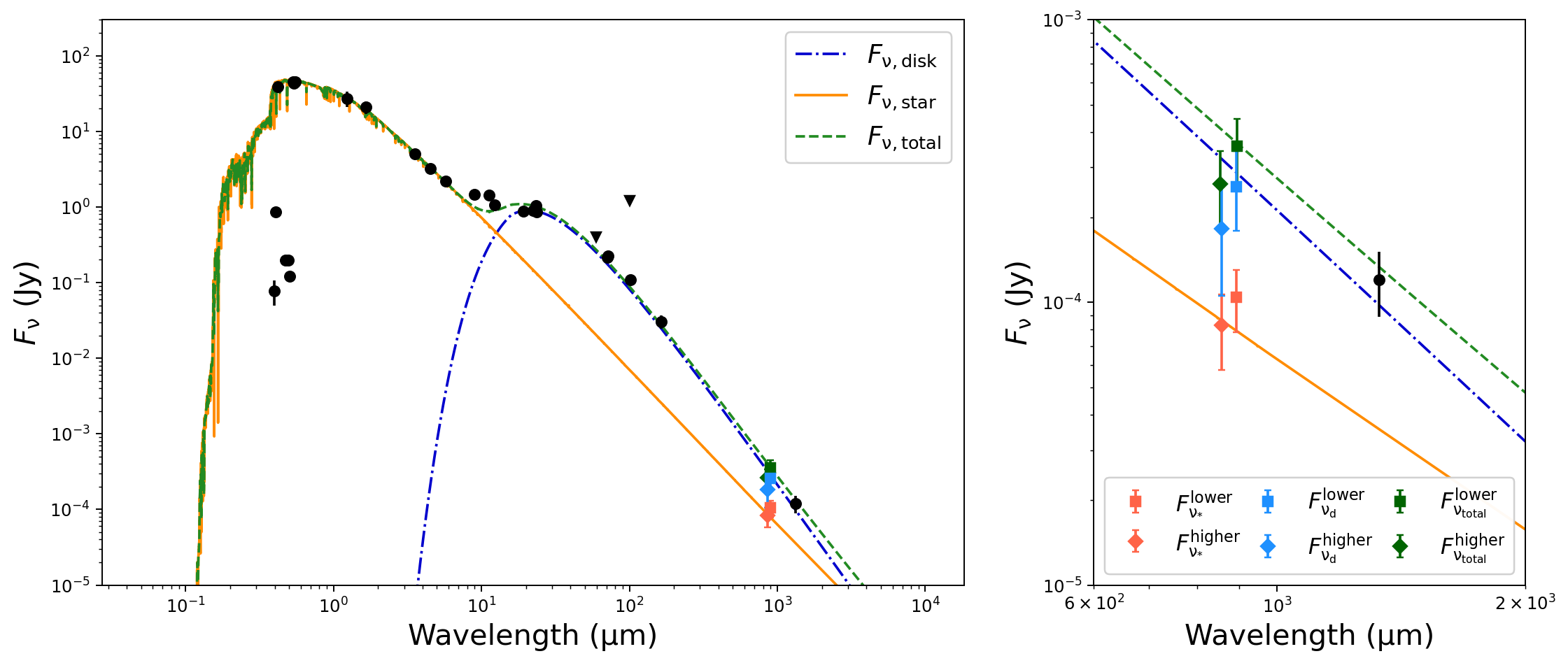}
    \caption{Left: full multi-wavelength photometry for HD 172555, including new millimeter photometry (colored symbols) from our observations. Right: zoom in from the left panel focusing on ALMA (sub)millimeter data. In both panels, diamonds correspond to the higher spectral setup and squares correspond to the lower spectral setup. Dark orange markers are stellar fluxes, blue markers are disk fluxes, and dark green markers are total (disk and stellar) combined fluxes. Error bars are $1\sigma$. Previous photometric measurements are in black, where circles are detections and downward-facing triangles are upper limits (photometric points are listed in Table \ref{tab:photometry}). The best-fit star (orange, solid) and single-component modified blackbody disk models (blue, dash-dotted) are included and obtained using the method from \citet{yelverton_statistically_2019}.}
    \label{fig:SED two panel}
\end{figure*}

 We create a combined disk and star spectrum (hereafter SED) from available multi-wavelength photometry and fitted it with a stellar model plus a modified blackbody model using the methodology from \citet{yelverton_statistically_2019}, shown in Fig. \ref{fig:SED two panel}. Our ALMA flux densities in each spectral setup are within 2$\sigma$ of the expected values from the SED, so our values are consistent with existing far-infrared to millimeter photometry, which constrains the millimeter spectral slope to $2.74 \pm 0.03$.
 
\par We also use this spectral slope to compare the HD 172555 disk to other debris disks and determine the particle size distribution. The millimeter spectral slope of debris disks at tens of au, assumed to have a steady state grain size distribution, typically range from 2.5-3 \citep[][and references therein]{hughes_debris_2018}. Our millimeter spectral slope of $2.74 \pm 0.03$ for HD 172555 is therefore consistent with the typical range for debris disks. The spectral slope can be linked to the slope of the grain size distribution through 
\begin{equation}
    q=\frac{\alpha_{\rm mm}-\alpha_{\rm Pl}}{\beta_{\rm s}}+3,
    \label{eqn:q}
\end{equation}
where $q$ is the power law index of the grain size distribution, $dn/da \propto a^{-q}$, $\alpha_{\rm mm}$ is the millimeter slope of the SED, $\alpha_{\rm Pl}$ is the spectral index of the Planck function between two frequencies, and $\beta_{\rm s}$ is the dust opacity spectral index of small particles, which is taken to be $1.8 \pm 0.2$ for astronomical silicates. The full derivation of Eq. \ref{eqn:q}, which assumes that the disk is dominated by grains smaller than the wavelength of the observation, uses an analytical formula derived by \citet{draineSubmillimeterOpacityProtoplanetary2006} and is applied in the context of debris disks in, e.g., \citet{ricciFomalhautDebrisDisk2012} and \citet{macgregorConstraintsPlanetesimalCollision2016}. Using Eq. \ref{eqn:q}, we find $q=3.41\pm 0.05$, which is consistent with steady state collisional models \citep[e.g.][]{dohnanyi_collisional_1969, macgregorConstraintsPlanetesimalCollision2016,  marshall_constraintsonmmemission_2017}. Interestingly, the size distribution slope measured at millimeter wavelengths is flatter than that found by \citet{lisse_abundant_2009} for smaller grains probed by infrared observations, $dn/da \propto a^{-3.95 \pm 0.10}$. This indicates a break in the size distribution, and the unique overabundance of small grains pointed out by \citet{lisse_abundant_2009}. \citet{johnson_self-consistent_2012} find that the emission observed by \citet{lisse_abundant_2009} is dominated by fine dust grains both smaller and larger than the blowout grain size of the system, so they will not be blown out of the system by radiation pressure. The grains smaller than the blowout grain size could have been created by the condensation of vaporized molten material from the initial impact \citep{lisse_abundant_2009}. The fine dust could be further created or replenished by subsequent collisions between debris from the initial impact \citep{johnson_self-consistent_2012}.

\section{Discussion}
\label{sect:discussion}
\subsection{Comparison to previous observations}
\label{sect:comp to previous obs}
\par In the millimeter, \citet{schneiderman_carbon_2021} detect the dust continuum and the \textsuperscript{12}CO J=2-1 transition, neither of which are spatially resolved. The \textsuperscript{12}CO is spectrally resolved and best described by a narrow ring of radius $7.4^{+0.5}_{-0.4}$ au and width $3.4\pm0.5$ au. The dust is constrained to be within 15 au of the star and is consistent with being co-located with the \textsuperscript{12}CO, both of which are consistent with our findings. \citet{schneiderman_carbon_2021} find a millimeter dust mass of (1.8$\pm$0.6)$\times 10^{-4}$ M$_\oplus$ from the observed continuum emission, assuming an equilibrium blackbody temperature of 169 K at $\sim$7.5 au and a dust grain opacity of 10 cm\textsuperscript{2} g\textsuperscript{-1} at 1000 GHz and scaling the dust grain opacity as $\kappa \propto \nu^\beta$ with $\beta=1$ \citep{beckwith_survey_1990}. For the 0.87 mm grains, we calculate a dust mass using the dust mass equation from \citet{hildebrand_determination_1983}, 
\begin{equation}
    M_{\rm dust} = \frac{F_{\rm \nu}d^2}{\kappa_{\rm \nu} B_{\rm \nu}(T)},
    \label{eqn:dust mass}
\end{equation}
where $M_{\rm dust}$ is the mass of the dust grains of a certain size in kg, $F_{\rm \nu}$ is the flux density of the dust grains in W m$^{-1}$ Hz$^{-1}$, $d$ is the distance to HD 172555 in pc, $\kappa_{\rm \nu}$ is the opacity of the dust grains at the observation frequency in cm$^2$ g$^{-1}$, and $B_{\rm \nu}(T)$ is the Planck function at the frequency of the observation as a function of the temperature of the disk. Unlike the grains dominating the SED, we assume the millimeter grains we observe act as blackbodies. Using an equilibrium blackbody temperature of 213 K at 4.7 au, the median peak surface density radius found in the Gaussian model, and scaling the dust grain opacity as described above, we calculate a dust mass of (9.4$\pm$1.1)$\times 10^{-5}$ M$_\oplus$. If we instead use an equilibrium blackbody temperature of 191 K at 5.9 au, the mean of our best-fit (mode) values of $R_{\rm in}$ and $R_{\rm out}$ from the power law model, we calculate a dust mass of (10.5$\pm$1.2)$\times 10^{-5}$ M$_\oplus$, which is consistent with the dust mass calculated from Gaussian model parameters. Our dust mass is lower than the dust mass calculated in \citet{schneiderman_carbon_2021} due to the higher temperature and smaller grains, and therefore lower opacities, considered here. 
\par In the far-infrared, in addition to atomic oxygen, \citet{riviere-marichalar_hd_2012} detect excesses from circumstellar dust at 70, 100, and 160 µm. \citet{riviere-marichalar_hd_2012} fit the excess with a modified blackbody model, yielding a temperature of 280$\pm$9 K, corresponding to a minimum blackbody dust radius of 2.7$\pm$0.2 au, which is in agreement with the location of the peak surface density in our observations within 1$\sigma$. 
\par In the mid-infrared, \citet{smith_resolving_2012} reanalyze imaging taken with the Thermal-Region Camera Spectrograph (TReCS) at the Gemini South telescope, originally presented by \citet{moerchen_high_2010}, to find that the warm dust emitting at $\sim$18 µm is in a disk with a median radius of 0.27" (7.8 au) from the central star, and they model the visibilities from interferometric data taken with the MID-infrared Interferometric instrument (MIDI) on the Very Large Telescope Interferometer to place a lower limit of 1 au on the radial location of the emitting region. They do not detect extended emission in the TReCS imaging data at $\sim$11 µm, allowing them to place an upper limit of 0.27" (7.8 au) on the spatial extent of the $\sim$11 µm emission. Because the $\sim$11 µm and $\sim$18 µm emission are at different locations, this indicates that this disk is extended, and the different bands of observation probe different regions of that disk due to their sensitivities to different temperatures \citep{smith_resolving_2012}. Also in the mid-infrared, \citet{samland_minds_2025} detect atomic gas lines $<$0.5 au from the central star and note that the lines appear to trace evaporating dust, asteroids, or exocomets. The detected gas disk is much closer to the central star than the dust we detect. \citet{samland_minds_2025} explore several options for the origin of the gas, including an evaporating population of close-in bodies, outgassing from a large in-situ rocky planet, or the inward drift and sublimation of fine dust. It is therefore possible that dust produced in the impact could be responsible for the observed atomic gas emission \citep{samland_minds_2025, su_mid-infrared_2020}.
\par Scattered light observations (using a filter covering the R and I bands, with $\lambda_{\rm c}=735$ nm and $\Delta\lambda=290$ nm, \citeauthor{engler_detection_2018} \citeyear{engler_detection_2018}) find the data are consistent with an axisymmetric dust disk with an outer radius between 0.3" and 0.4" (8.6-11.5 au). They do not detect an inner edge to the disk, but note their observations do not probe interior to 5 au from the star, and their data allow the presence of compact emission closer to the star, behind the occulting spot of the coronagraph \citep{engler_detection_2018}. 
\par The findings from both scattered light and the mid-infrared are consistent with our millimeter observations. \citet{smith_resolving_2012} report that a disk position angle PA of 120º and an inclination $i$ of 75º best fit their data, and place the limits that 40º $<$ PA $<$ 130º and $i >$47º. \citet{engler_detection_2018} find a position angle of 112.3º$\pm$1.5º and inclination of 76.2º$\pm$1.7º best fit their data. The radial location and geometry of the disk in marginally resolved scattered light and mid-infrared observations therefore indicate that small grains, as well as large millimeter-sized grains, are present in the same $\lesssim$11.5 au region from the central star.

\subsection{Radial distribution}
\label{sect:radial distribution}
\par The results of our Gaussian model (Table \ref{tab:final values}, middle column) show that the disk is likely relatively broad, though not well constrained due to the S/N of our observations. In particular, our data excludes Gaussian models that are jointly compact (small $R$) and narrow (see joint R, $\Delta R$ probability distribution in Fig. \ref{fig:cornerplot incl gaussian prior 10000 steps}) because emission is clearly resolved over several resolution elements in our images. Additionally, peak surface density radii larger than $\sim$10 au are excluded because the detected emission only extends to $\sim$8.6 au. Overall, if we assume the Gaussian model, our data indicate that the surface mass density distribution is likely broader than a few au, with broader models peaking closer to the central star, and with no lower limit on a peak surface mass density distribution radius location, as emission is detected all the way to the central resolution element (Fig. \ref{fig:all data plus diskonly}). Even broader models beyond $\sim$8.6 au cannot be excluded if they are lower in surface brightness and therefore undetected at that boundary. 

\par The results of our power law model (Table \ref{tab:final values}, right column) show the disk is more likely to be relatively narrow, although, again, not well constrained due to the S/N of our observations. $R_{\rm in}$ is not constrained, although there is a marginally higher probability of $R_{\rm in}$ being larger, $\sim$5 au, rather than smaller ($<$3 au). While $R_{\rm in}$, $R_{\rm out}$ pairs that span the range of the priors are allowed in the power law model, there is a preference for narrower disks, as the joint $R_{\rm in}$, $R_{\rm out}$ probability distribution peaks at larger $R_{\rm in}$, $\sim$5 au, and smaller $R_{\rm out}$, $\sim$7 au (see Fig. \ref{fig:cornerplot incl gaussian prior 10000 steps power law}). Overall, if we assume the power law model, our data indicate that the surface mass density distribution is likely narrower and only a few au wide. 

\par Our modeling results show that determining the radial extent of the HD 172555 dust disk is highly dependent on the model for the radial surface mass density distribution, highlighting the need for higher S/N observations. We note that parameters describing the radial surface density distribution of material are model-dependent; the rest (fluxes, $i$, PA, $\Delta$RA, $\Delta$Dec, and $f$) are robust to changing model assumptions.

\par From mid-infrared imaging, \citet{smith_resolving_2012} find that the observed emission can be fit by disks with radii from 0.09"-0.31" (2.6-8.9 au) and widths between 0.36$r$ and 2$r$, where $r$ is the radius of the disk, implying that the grains probed by the $\sim$18 µm imaging have a broad distribution, which is supported by our Gaussian model results. Additionally, \citet{lisse_abundant_2009} fit the excess dust spectrum and model the grain composition to find that the small grains they observe, including the ones creating the silica feature at $\sim$9 µm, are 5.8$\pm$0.6 au from the central star. The location of the small grains from \citet{lisse_abundant_2009} produced in the impact are then consistent with the location of the millimeter grains. 
\par
\begin{figure}
    \centering
    \includegraphics[width=\hsize]{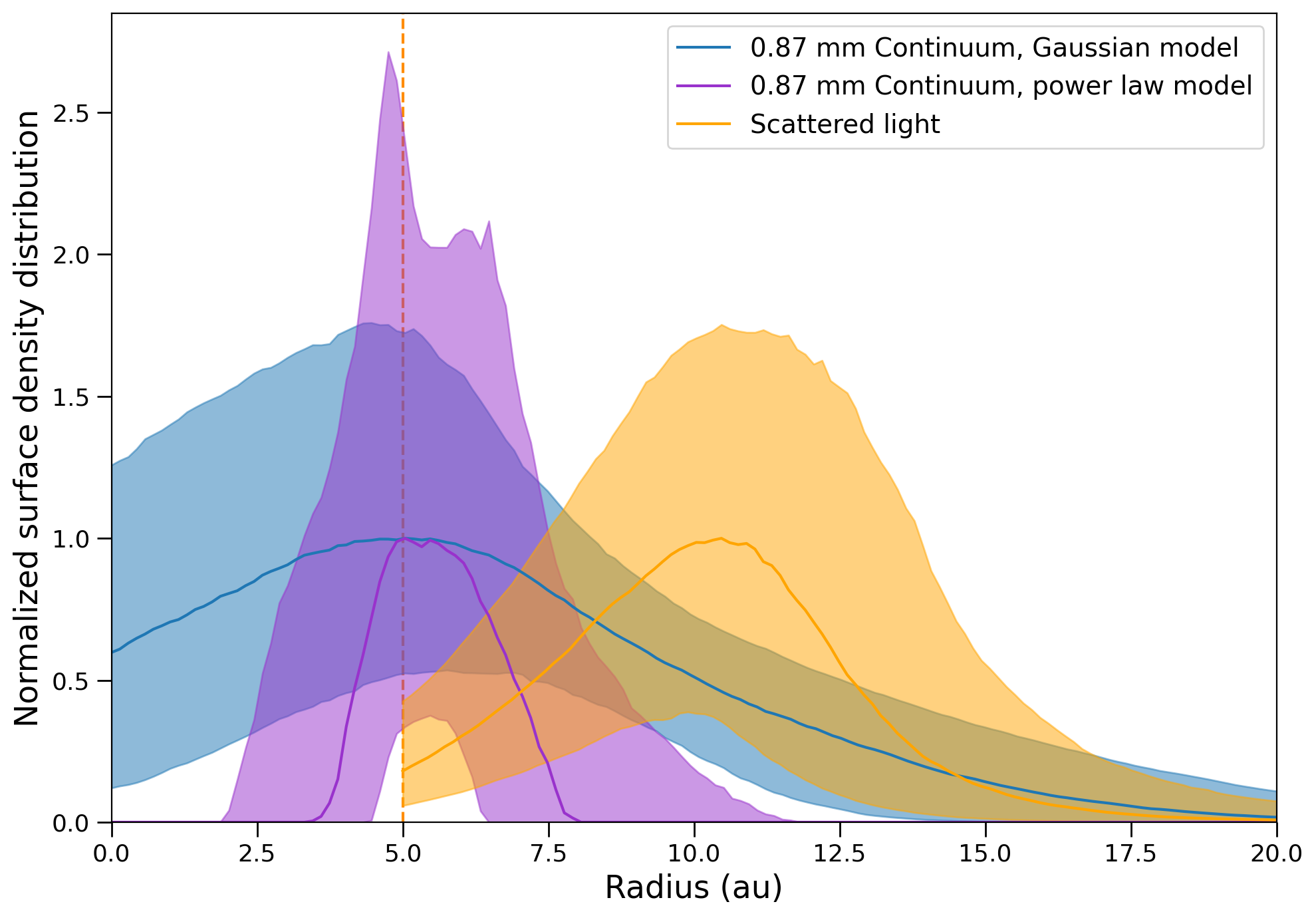}
    \caption{Surface density distribution of the millimeter grains from this paper (Gaussian model in blue, power law model in purple) compared to the the surface density distribution from scattered light imaging (orange, \citeauthor{engler_detection_2018} \citeyear{engler_detection_2018}). The orange dashed line is the inner limit of the scattered light observations. All curves have been normalized to 1, and the shaded regions correspond to 1$\sigma$ error.}
    \label{fig:dust comp to co}
\end{figure}

 In Fig. \ref{fig:dust comp to co}, the blue curve shows the normalized surface mass density, in arbitrary units, calculated from randomly sampled $R$, $\Delta R$ pairs from the Gaussian model MCMC samples. This curve shows the likely broad distribution of the millimeter dust in the HD 172555 disk the Gaussian model finds. The purple curve is the same as the blue one, but calculated using randomly sampled $R_{\rm in}$, $R_{\rm out}$, $p$ triples from the power law model MCMC samples. The orange curve in Fig. \ref{fig:dust comp to co} is a scaled surface mass density obtained using the best-fit disk geometry and modeling parameters from the scattered light observations, where we converted from number density given by \citet{engler_detection_2018} to surface mass density using their assumed vertical density structure. From these, it is clear that distributions of the millimeter grains and the smaller ones probed by the scattered light observations are offset, where the millimeter grains are closer to the star than the smaller grains, potentially by a factor of two, although there is a high uncertainty on this value. 
\par Given the known presence of gas, the difference in the radial location of the grains could be due to radiation pressure with gas drag \citep[e.g.][]{takeuchi_dust_2001}. Both radiation pressure from the central star and gas drag affect grains of different sizes differently, causing them to potentially migrate inward or outward until the grains feel no net torque \citep{takeuchi_dust_2001}. For radiation, it is helpful to compare how the forces of radiation and gravity affect a grain's orbit. The ratio between radiation and gravity is $\beta$, which, from \citet{burns_radiation_1979}, is given by
\begin{equation}
    \beta=\frac{3L_*Q_{\rm PR}}{16 \pi cGM_*a\rho_{\rm d}},
    \label{eqn:beta}
\end{equation}
where $L_*$ is the luminosity of the central star, $Q_{\rm PR}$ is the radiation pressure coefficient averaged over the stellar spectrum, $M_*$ is the mass of the central star, $a$ is the radius of a grain, and $\rho_{\rm d}$ is the material density of the grain, all in SI units. From Eq. \ref{eqn:beta}, it is clear smaller grains will feel a stronger force from the stellar radiation than larger grains and will therefore be pushed onto more eccentric orbits, creating a dust distribution where smaller grains are seen at larger orbital distances, assuming other frictional forces are negligible \citep{krivov_debris_2010}. 
\par In the presence of gas, a standard way to characterize the behavior of solids is with the Stokes number, which quantifies the extent to which a grain is affected by gas drag by comparing the stopping timescale $t_{\rm s}$ to the orbital timescale $\Omega_{\rm k}^{-1}$ \citep[e.g.][]{birnstiel_gas-_2010}. We express the Stokes number, St, as
\begin{equation}
    {\rm St}=\frac{\rho_{\rm d}}{\rho_{\rm g}}\frac{a}{v_{\rm th}}\Omega_{\rm k},
    \label{eqn:stokes}
\end{equation}
where $\rho_{\rm g}$ is the density of the gas, and $v_{\rm th}$ is the thermal velocity of the gas. 
\par It is then possible for the combined effect of gas drag and radiation pressure from the central star, if the gas density $\rho_{\rm g}$ is high enough, to cause outward drift of the smaller grains without affecting the millimeter grains, if the Stokes numbers of each grain are right for those conditions. Without knowing the gas density, and therefore composition beyond CO, it is difficult to accurately determine the Stokes numbers for the grain sizes probed by observations. Assuming the gas is dominated by CO as detected by \citet{schneiderman_carbon_2021}, we take both the optically thick ($\sim$38 K) and optically thin (169 K) CO masses they derived and find Stokes numbers in the range of $10^{-2}-10^3$ for micron-sized grains and $10^2-10^6$ for millimeter-sized grains. It is therefore plausible, for the larger masses implied if \textsuperscript{12}CO is optically thick, or if other species dominate the gas mass, that the micron-sized grains drift outward while the millimeter grains would be unaffected. However, full radiative transfer gas modeling is needed for quantitative interpretation of the gas drag, as, for example, the mass and radial distribution of the \textsuperscript{12}CO detected by \citet{schneiderman_carbon_2021} are degenerate and model-dependent, respectively.

\subsection{Mass of the largest impacting planet}
\label{sect:mass of largest planet}
\par Using the radial width of the disk of millimeter-sized grains, we can constrain the mass of the largest impacting planet, assuming gas has not impacted the dynamics of the millimeter and larger sized grains throughout the post-impact evolution. Following from \citet{jackson_debris_2014}, the radial width of the debris from a giant impact is directly related to the velocity dispersion of the debris. The debris is expected to be ejected from the impacting planetary embryos with relative velocities of order the escape velocities of the impactors, which makes the radial width of the grains the best proxy for progenitor mass. A more massive progenitor would have a higher escape velocity, which would result in a broader distribution of millimeter-sized grains compared to a less massive progenitor. From \citet{schneiderman_carbon_2021}, the expected mass of the impacting planet in M$_\oplus$ is 

\begin{equation}
M_{\rm p} \sim 103 \frac{M_*^{3/2}}{\rho_{\rm p}^{1/2}} (\frac{\Delta R}{R^{3/2}})^{3},
\label{eqn:planet mass}
\end{equation}

where $M_{\rm *}$ is the stellar mass in M$_\odot$, $\rho_{\rm p}$ is the bulk density of the planet in g cm$^{-3}$, and $R$ and $\Delta R$ are in au. Using $M_*$ = 1.76M$_\odot$, $\rho_{\rm p}$ = 5.5 g cm$^{-3}$, and randomly sampled $R$, $\Delta R$ pairs from the Gaussian model MCMC samples, we derive a mass estimate of $0.5^{+19.2}_{-0.4}$ M$_{\rm Jup}$. However, this estimate is not well constrained due to the high uncertainties on the millimeter dust radius and width from the Gaussian model (see Sect. \ref{sect:radial distribution}). Additionally, a planet with this mass is likely to be a gas giant, which means that the outcome of a collision with such a planet is unlikely to be large amounts of dust, as we assumed. 
\par We can also use the results from the power law model to estimate the mass of the largest impacting planet. We randomly sample $R_{\rm in}$, $R_{\rm out}$ pairs from the power law model MCMC. In Eq. \ref{eqn:planet mass}, we take $R$ to be the mean of the  $R_{\rm in}$ and $R_{\rm out}$ and $\Delta R$ to be the difference between $R_{\rm out}$ and $R_{\rm in}$. Again using $M_*$ = 1.76M$_\odot$ and $\rho_{\rm p}=5.5$ g cm$^{-3}$, we derive an estimate of $4.5^{+13.5}_{-3.3}$ M$_{\oplus}$ ($0.014^{+0.042}_{-0.010}$ M$_{\rm Jup}$) for the mass of the largest impacting planet. A planet of this mass is likely to be a terrestrial planet, meaning the outcome of a collision with this planet would be consistent with our assumptions about the debris from the giant impact. While we cannot distinguish between the two models with our data, the mass of the largest impacting planet we derive using the Gaussian model is unlikely, while the mass derived using the power law model is more consistent with our assumptions. This makes the power law model physically preferable to the Gaussian model.

\par Although it is unclear how massive the largest impactor was, we can consider timescales for debris survival in the disk as it encounters a leftover planet \citep{wyattHowDesignPlanetary2017}. Conservatively, we assume that a leftover planet will have a mass of order the mass of the largest impactor. Using Eq. 2 from \citet{brasser2008}, with a semimajor axis of 4.7 au, the eventual outcome for debris encountering a planet of 0.5 M$_{\rm Jup}$ is ejection from the system, with an ejection timescale of $\sim$0.8 Myr. A less massive remaining planet would have a longer ejection timescale, and for planet masses below a certain threshold, the predominant outcome for debris encounters will be accretion. The eventual outcome for debris encountering a 4.5 M$_\oplus$ leftover planet at $\sim$5 au is accretion onto the remaining planet, although the accretion timescale is much longer than the age of the system. It remains possible that the mass of the largest impactor was $\sim$0.5 M$_{\rm Jup}$, as long as we are observing this system within $\sim$0.8 Myr of the impact. It is also possible that the mass of the largest impacting planet was 4.5 M$_\oplus$, as debris would survive encounters with a leftover planet of order 4.5 M$_\oplus$ for longer than the age of the system. 
\par We can also consider the limits on the presence of any planets in the system that may affect post-impact debris evolution. \citet{quanz_searching_2011} used VLT/NACO to search for planetary mass companions to HD 172555 at 4 µm, and while they did not detect any planets, they derived detection limits of 2-3 M$_{\rm Jup}$ at projected separations of 15-29 au and $\gtrsim$4 M$_{\rm Jup}$ at $\sim$11 au. \citet{meunier_comparison_2012} used radial velocity data to place detection limits of between $\sim$4 M$_{\rm Jup}$ for short period orbits (0.6 au) and $\sim$30-70 M$_{\rm Jup}$ for long period orbits (2.1 au). 
\par While our derived impactor masses are within these observational limits, and while debris could survive repeat encounters with the more massive planet estimate up to $\sim$0.8 Myr post-impact, the currently large uncertainties, particularly on the mass derived using the results from the Gaussian model, prevent us from drawing firm conclusions. Deeper observations of gravitationally bound grains will be needed to better constrain the radial width of the debris and, by extension, the mass of the largest impacting planet.

\subsection{Limits on potential asymmetry}
\label{sect:time since impact}
\par Motivated by marginal evidence for an asymmetry in the dust distribution detected by \citet{smith_resolving_2012} and \citet{engler_detection_2018}, with the southeast side being brighter, we attempt to constrain the magnitude of an asymmetry in the millimeter by assuming that we would just be able to detect an asymmetry if the ratio of the peak intensity on each side of the disk was inconsistent with 1 by at least 3$\sigma$. Using the maximum intensity of (5.1$\pm$0.9)$\times$10$^{-5}$ Jy beam$^{-1}$ on the southeast side of the disk and (4.2$\pm$0.9)$\times$10$^{-5}$ Jy beam$^{-1}$ on the northwest side of the disk in the right panel of Fig. \ref{fig:all data plus diskonly}, we find that the ratio of the southeast intensity, $I_{\mathrm{\nu, SE}}$, to the northwest intensity, $I_{\mathrm{\nu, NW}}$,  is 1.21$\pm$0.39, which is consistent with 1, implying that no southeast-northwest asymmetry is detected in our data. Detection would have only been achieved if $\frac{I_{\mathrm{\nu, SE}}}{I_{\mathrm{\nu, NW}}}$ were $3\sigma$ above 1, or $\frac{I_{\mathrm{\nu, SE}}}{I_{\mathrm{\nu, NW}}}=2.17$. We therefore constrain a possible asymmetry in the millimeter dust distribution to not more than a 117\% difference between the southeast and northwest sides of the disk, assuming the asymmetry is in the plane of the sky. It is then plausible for a weak to moderate asymmetry to still be present for the millimeter grains and compatible with the marginal asymmetry seen in shorter wavelength observations \citep{engler_detection_2018, smith_resolving_2012}.

\par Whether or not an asymmetry is present would allow us to constrain the time that has passed since the impact. The debris from a giant impact is produced at one point in space, which means that, in the absence of sufficient amounts of gas to affect the grain dynamics, all of the debris orbits start at this location and must pass through this point again \citep{jackson_debris_2014}. Because all of the debris orbits must then pass through the impact point, the models show that this creates a pinch point in the disk with enhanced dust production from further collisions between debris produced in the impact, leading to an azimuthal asymmetry in the dust structure \citep{jackson_debris_2014}. The asymmetry then smears out over time as the orbits precess from interactions with leftover planets, other planets in the system, or the post-impact disk's own self-gravity, to eventually create an axisymmetric disk. The timescale for a disk to become centrally symmetric depends strongly on the semimajor axis of the debris, as well as on the semimajor axis and mass of planets with which the debris interacts, and could be of order a few tens of thousands of debris orbits \citep{jackson_debris_2014, kral2015}. In the HD 172555 system, this would translate to a timescale of $\sim$0.3 Myr for the disk to become symmetric. It is possible that we are observing this system $\gtrsim$0.3 Myr post-impact, especially given the accretion timescale discussed in Sect. \ref{sect:mass of largest planet} if the mass of the largest impacting planet was of order 4.5 M$_{\oplus}$. In the high impactor mass scenario, given the debris ejection timescale discussed in Sect. \ref{sect:mass of largest planet}, it is possible (although unlikely) that we are observing this system between $\sim$0.3 and $\sim$0.8 Myr post-impact. We cannot accurately constrain the time that has passed since the impact until the disk's radial width and level of asymmetry are more accurately constrained with higher S/N observations. 

\section{Summary}
\label{sect:summary}
\par In this work, we present the first spatially resolved 0.87 mm continuum observations of post-impact dust around HD 172555 obtained with ALMA. We imaged and modeled the interferometric visibilities of our data to analyze the disk structure and report the following findings:
\begin{itemize}
    \item We clearly detect the star and the disk; after subtracting the star interferometrically, we reveal an inclined disk of millimeter dust emission extending out to $\sim$9 au, and down to within 2.3 au of the star on-sky.
    \item We measure millimeter fluxes implying a total dust mass of $(9.4\pm1.1)\times 10^{-5}$, and include the fluxes in SED modeling, which reveals a spectral slope of $2.74 \pm 0.03$. This implies a size distribution slope of 3.41$\pm$0.05, which differs from the size distribution measured for micron-sized grains and implies that there is a break in the grain size distribution shortward of millimeter-sized grains. 
    \item Modeling the visibilities indicates that the radial disk surface density distribution most likely peaks around $\sim$5 au, while the radial width inferred remains dependent on the model. Our Gaussian model is more likely a broad disk than a narrow ring, while our power law model is more likely to be a narrow ring between $\sim$5 and $\sim$7 au. 
    \item We report a radial offset between the millimeter grains and the small grains traced by scattered light observations (peaking at $\sim11$ au). This could be due to the combined effect of radiation pressure and gas drag, if the gas density is high enough.
    \item Assuming our best-fit values of disk radius and width (Gaussian model) or inner and outer radius (power law model) leads to estimates of the largest impacting planet mass of 0.5 $M_{\rm Jup}$ and 4.5 $M_\oplus$, respectively. In the large planet case, the impact is unlikely to be dust dominated, as has been assumed, and had to be very recent, as the ejection timescale is very short, whereas in the smaller, terrestrial planet case, most of the debris could have avoided dynamical removal even if the impact happened very early in the system's lifetime. However, our estimates remain highly uncertain with the S/N of our observations.
    \item We do not see significant evidence of an asymmetry in the dust distribution, although a strong asymmetry of $<$117\% between the southeast and northwest sides of the disk could still be present and have gone undetected. 
\end{itemize}

\par Further work is needed to better understand the post-impact dynamics in this system. Higher S/N observations would allow us to better constrain the radial distribution of the gravitationally bound dust, which would in turn lead to tighter limits on the mass of the largest impactor and the time that has passed since the impact. Additionally, to ascertain whether the gas in the disk impacts the dust dynamics, higher S/N ALMA and scattered light dust observations are needed to more accurately constrain a radial offset between the two. In parallel, an upcoming in-depth analysis of the gas composition and distribution using this ALMA data will allow us to constrain the disk's volatile content and dynamics, shedding light on the impacting bodies and the subsequent debris evolution.

\begin{acknowledgements}
\par ZR and LM acknowledge funding by the European Union through the E-BEANS ERC project (grant number 100117693). SM acknowledges funding by the Royal Society through a Royal Society University Research Fellowship (URF-R1-221669) and the European Union through the FEED ERC project (grant number 101162711). Views and opinions expressed are however those of the author(s) only and do not necessarily reflect those of the European Union or the European Research Council Executive Agency. Neither the European Union nor the granting authority can be held responsible for them.
\par This paper makes use of the following ALMA data: ADS/JAO.ALMA\#2022.1.00793.S. ALMA is a partnership of ESO (representing its member states), NSF (USA) and NINS (Japan), together with NRC (Canada), NSC and ASIAA (Taiwan), and KASI (Republic of Korea), in cooperation with the Republic of Chile. The Joint ALMA Observatory is operated by ESO, AUI/NRAO and NAOJ. 

\end{acknowledgements}

\bibliographystyle{aa}
\bibliography{references}

\begin{appendix}
\onecolumn
\section{MCMC results}
\label{sect:appendix mcmc results}
\begin{figure*}[h!]
    \centering
    \includegraphics[width=\hsize]{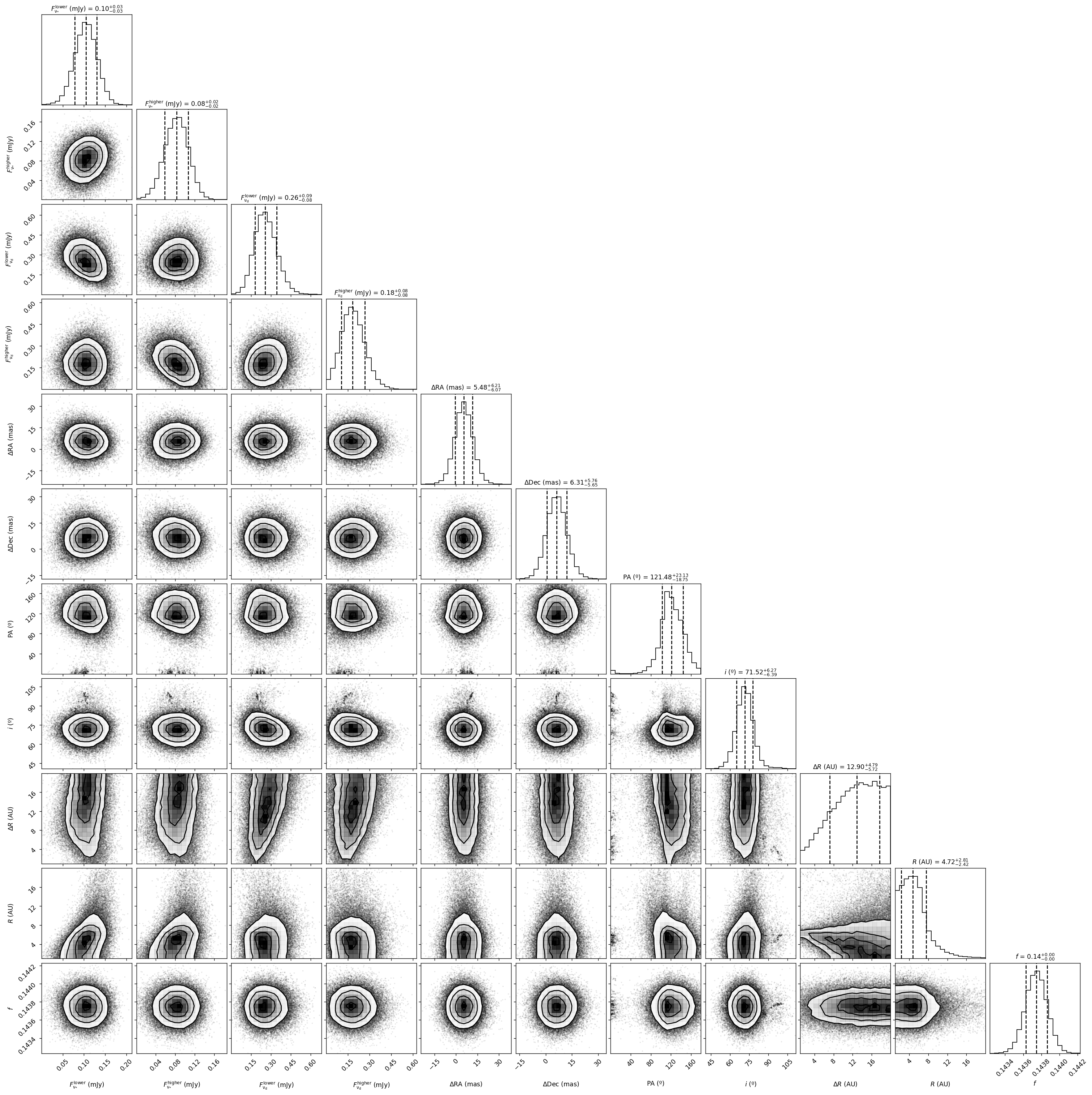}
    \caption{One- and two-dimensional posterior probability distributions for all of the MCMC parameters from the Gaussian model (described in Sect. \ref{sect:physical model}), where we applied a Gaussian prior to the inclination $i$. Listed above each one-dimensional posterior probability distribution is the $50^{+34}_{-34}$th percentile value of the posterior probability distribution of each parameter marginalized over all other parameters.}
    \label{fig:cornerplot incl gaussian prior 10000 steps}
\end{figure*}

\begin{figure*}[h!]
    \centering
    \includegraphics[width=\hsize]{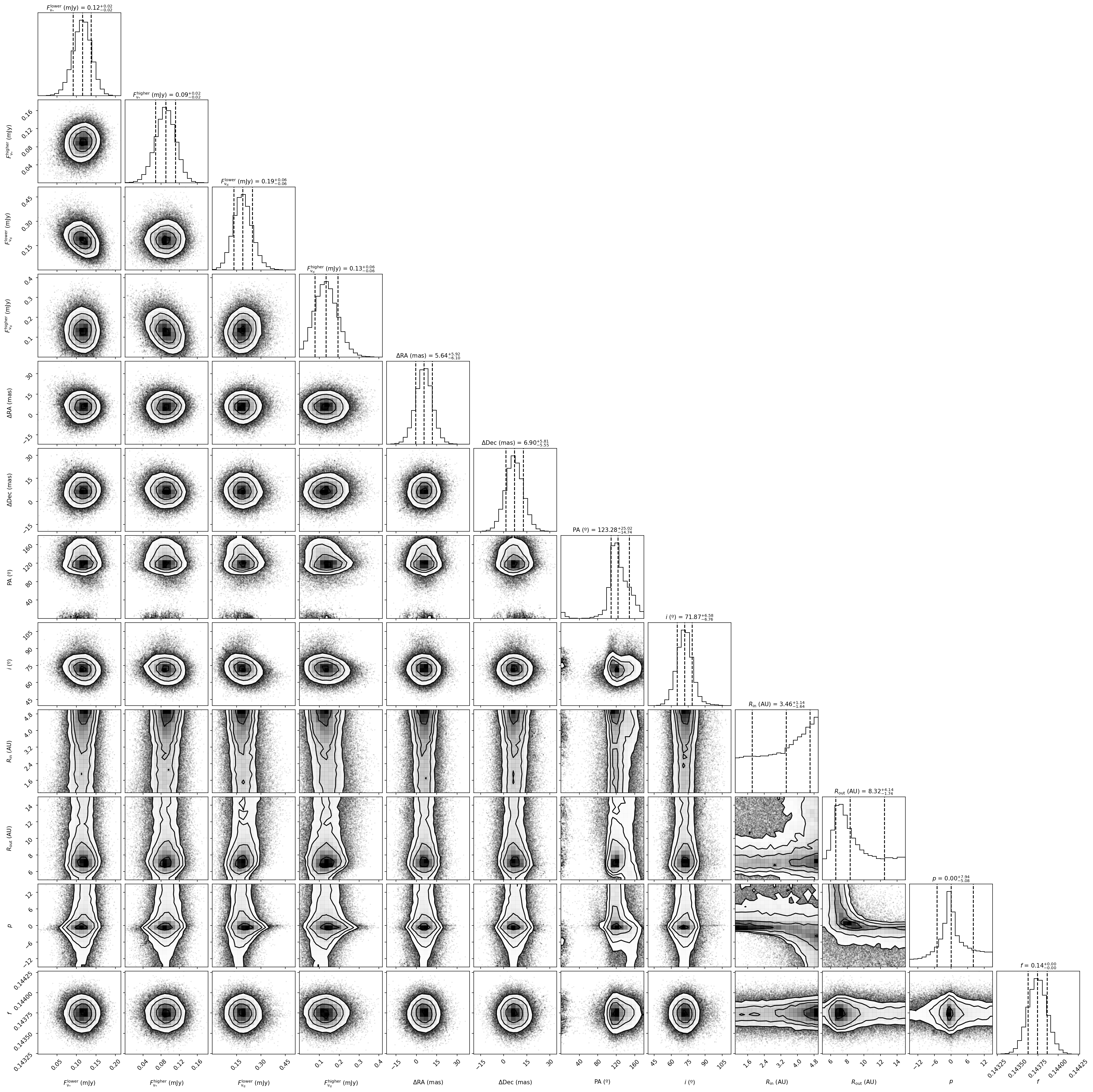}
    \caption{Same as Fig. \ref{fig:cornerplot incl gaussian prior 10000 steps}, but for the power law model (described in Sect. \ref{sect:physical model}).}
    \label{fig:cornerplot incl gaussian prior 10000 steps power law}
\end{figure*}

\clearpage
\section{Photometry}
\label{sect:photometry origins}

\begin{table*}[!htb]
\caption{Photometric values}
\centering {
\begin{tabular}{lllllll}
$\lambda_{\rm c}$ & Magnitude & $\sigma_{\rm mag}$ & Flux Density & $\sigma_{\rm flux}$    & Ref.              \\
(µm) & (mag) &  (mag) & (Jy) & (Jy) & \\
\hline
0.40      & 0.08                  & 0.03     & --  & --  &  1   \\
0.41       & 0.859                  & 0.011     & --  & -- & 2 \\
0.42       & -- & -- & 38.9            & 0.5      & 3  \\
0.47       & 0.196                  & 0.015      & --  & -- & 2  \\
0.49       & 0.20                  & 0.02      & --  & --  & 1  \\
0.51       & 0.122                 & 0.012      & --  & --  & 2 \\
0.53       & -- & -- & 45.3        & 0.4   & 3 \\
0.54       & -- & -- & 43.1       & 0.3     & 4  \\
0.55       & -- & -- & 45.1        & 0.9     & 1  \\
1.24       & -- & -- & 27.3       & 6.5       & 5  \\
1.65       & -- & -- & 20.9     & 4.1        & 5  \\
3.56       & -- & -- & 5.09    & 0.10  & 6         \\
4.51       & -- & -- & 3.20       & 0.06  & 6        \\
5.74       & -- & -- & 2.20        & 0.04  & 6         \\
8.98       & -- & -- & 1.45    & 0.03    & 7 \\
11.23       & -- & -- & 1.43    & 0.13       & 8  \\
12.33      & -- & -- & 1.06      & 0.05      & 9  \\
19.21       & -- & -- & 0.87     & 0.02    & 7 \\
22.25       & -- & -- & 0.91     & 0.05     & 9  \\
23.34       & -- & -- & 1.03      & 0.08        & 8  \\
23.68       & -- & -- & 0.87    & 0.02      & 10        \\
59.35        & -- & -- & $<$ 0.40 & 0.04          & 8  \\
71.15       & -- & -- & 0.214     & 0.006  & 11             \\
71.42       & -- & -- & 0.226     & 0.013     & 10        \\
100.35       & -- & -- & $<$ 1.20 & 0.12         & 8  \\
101.40       & -- & -- & 0.108   & 0.004   & 11             \\
163.60       & -- & -- & 0.031     & 0.006    & 11         \\
856.65       & -- & -- &   0.00026      &   0.00009        & 12 \\ 
893.03       & -- & -- &   0.00036     &  0.00009           & 12 \\
1322.43       & -- & -- & 0.00012    & 0.00003        & 13
\end{tabular}
}
\tablefoot{Upper limits are 3$\sigma$. Errors include systematic uncertainty. All photometric values include the disk as well as the central star.}
\tablebib{(1) \citet{mermilliod_vizier_2006} (2) \citet{paunzen_new_2015} (3) \citet{hog_tycho-2_2000} (4) \citet{esa_hipparcos_1997} (5) \citet{cutri_2mass_2003} (6) \citet{irsaSpitzerEnhancedImaging2020} (7) \citet{ishihara_akariirc_2010} (8) \citet{helou_infrared_1988} (9) \citet{wright_wide-field_2010} (10) \citet{gaspar_collisional_2013} (11) \citet{marton_herschelpacs_2017} (12) This work (13) \citet{schneiderman_carbon_2021}}
\label{tab:photometry}
\end{table*}

\end{appendix}

\end{document}